\documentclass[notitlepage,10pt,oneside,reqno]{amsart}
\usepackage[left=1in,right=1in,top=1in,bottom=1in]{geometry}
\usepackage{xcolor}
\usepackage[noadjust]{cite}
\numberwithin{equation}{section}
\newcommand{\lab}[1]{\bigg\langle{#1\bigg\rangle}}
\newcommand{\la}{\bigg\langle}
\newcommand{\ra}{\bigg\rangle}
\newcommand{\z}{\overline{z}}
\definecolor{lapis}{rgb}{0.0.0470,0.2941,0.5568}
\definecolor{burgundy}{rgb}{0.5, 0.0, 0.13}
\usepackage[colorlinks=true, linkcolor=burgundy, urlcolor=lapis, citecolor=lapis, anchorcolor=green]{hyperref}  
\title{\textsc{On Chiral Splitting and the Ambitwistor String}}
\author{\vspace{-0.5cm}Nikhil Kalyanapuram}
\address{{Department of Physics, Pennsylvania State University, University Park PA 16802, USA}}
\address{{Institute for Gravitation and the Cosmos, Pennsylvania State University, University Park, PA 16802, USA}}
\email{nkalyanapuram@psu.edu}

\usepackage{lipsum}
\begin{document}

\begin{abstract}
Scattering amplitudes computed by superstring perturbation theory are known to holomorphically split into chiral half integrands at fixed internal loop momentum. It is established by direct computation that upon reduction to the ordinary moduli space, the chiral half integrands of the superstring match those computed by the ambitwistor string in the limit of zero tension ($\alpha'\rightarrow\infty$). Subtleties that arise at higher genus due to the nonprojectedness of the supermoduli space are considered and arguments as to their resolution are furnished. 
\end{abstract}

\maketitle
\tableofcontents
\section{Introduction}
The use of string theory in the computation of scattering amplitudes in quantum field theory has a long history. For example, string-based rules were provided by Bern and Kosower \cite{Bern:1987tw,Bern:1990cu,Bern:1990ux,Bern:1991aq} to compute scattering amplitudes at one loop in QCD. A more formal development came from Witten's twistor string construction \cite{Witten:2003nn}, in which it was established that at tree level MHV amplitudes in $\mathcal{N}=4$ super Yang-Mills can be equivalently understood as a $D$-instanton expansion of the topological B-model on the Calabi-Yau supermanifold $\mathbb{CP}^{3|4}$. This was later generalised to other helicity sectors by Roiban, Spradlin and Volovich \cite{Roiban:2004vt,Roiban:2004yf}. The twistor string approach has since been generalised to a large class of quantum field theories in four and higher dimensions \cite{Cachazo:2012da,Cachazo:2012kg,Heydeman:2017yww,Cachazo:2018hqa,Heydeman:2018dje}.

An extension of the worldsheet picture of scattering amplitudes to theories with less (or no) supersymmetry was realised by the formalism of Cachazo, He and Yuan \cite{Cachazo:2013gna,Cachazo:2013hca,Cachazo:2013iaa,Cachazo:2014nsa,Cachazo:2014xea}. In this framework, scattering amplitudes in a vast array of quantum field theories (one that includes, notably, Yang-Mills and Einstein gravity) are recast as localised integrals over the moduli space $\mathcal{M}_{0,n}$ of marked Riemann spheres

\begin{equation}
    \mathcal{A} = \int_{M_{0,n}}d\mu_{n}\mathcal{I}_{L}\mathcal{I}_{R}\prod_{i}\delta\left(E_{i}\right).
\end{equation}
The conditions $E_{i}$, known as the scattering equations, take the form

\begin{equation}\label{eq:1.2}
    E_{i} = \sum_{j\neq i}\frac{k_{i}\cdot k_{i}}{z_{i}-z_{j}}
\end{equation}
where $z_{i}$ are marked points. These conditions had previously appeared in work by Gross and Mende \cite{Gross:1987ar} as saddle point constraints in the tensionless limit of superstring scattering amplitudes. The half integrands $\mathcal{I}_{L,R}$ encode kinematic data and are rational functions of the marked points. Judicious choices of these integrands lead to a wide variety of theories admitting a representation of the form (\ref{eq:1.2}).

A natural question to pose at this point is how one would generalise this framework to higher loop order. This problem was solved when it was observed that the half integrands in the CHY formalism can be equivalently obtained as correlation functions of a chiral string theory known as the ambitwistor string \cite{Mason:2013sva}. The ambitwistor string is a critical superstring with its target space as the space of null geodesics in Minkowski space. Unlike the conventional superstring, the ambitwistor string is purely holomorphic, and is built out of two sets of worldsheet fields. The ambitwistor string has been used to reproduce the CHY formulae for Yang-Mills and gravity at tree level \cite{Mason:2013sva} as well as to derive the appropriate generalisations at one loop \cite{Adamo:2013tsa,Geyer:2015bja,Geyer:2015jch} and two loop orders \cite{Geyer:2016wjx,Geyer:2018xwu}.

While there was heuristic evidence to suggest that the ambitwistor string was an infinite tension limit of the Ramond-Neveu-Schwarz superstring \cite{Mason:2013sva}, at least at the level of the worldsheet, a precise relation remained undiscovered. It was later seen to be the case that a more natural interpretation of the ambitwistor string is as a tensionless limit of the superstring \cite{Ohmori:2015sha,Casali:2016atr}. 

In this paper, we make use of the chiral splitting theorem due to D'Hoker and Phong \cite{DHoker:1989cxq} to prove by direct evaluation that the holomorphic chiral integrand in superstring theory reduces to the corresponding half integrand of the ambitwistor string in the limit of zero tension. Making this precise, we show that the chiral integrand for $n$ NS states at genus $g$ in the RNS string at fixed spin structure $\delta$ takes the form

\begin{equation}
    A_{g,n}[\delta] = \mathrm{RN}\times \mathcal{I}(\alpha')
\end{equation}
where $\mathrm{KN}$ provides a generalisation of the (holomorphic) Koba-Nielsen factor at genus $g$ and $\mathcal{I}(\alpha')$ upon taking the limit $\alpha'\rightarrow \infty$ tends to the chiral integrand computed by the ambitwistor string.
\vspace{1cm}
\paragraph{\textbf{Outline.}} 
In section \ref{sec:2} the basic facts of string perturbation theory on supermanifolds are reviewed, followed by a statement of the chiral splitting theorem. The holomorphically split amplitudes are then defined as correlation functions. The actual computation of the chiral integrand is carried out in full generality in section \ref{sec:3}. This is followed by section \ref{sec:4} in which the local reduction of the chiral integrand to $\mathcal{M}_{g,n}$ is performed and the accompanying subtleties and complications are considered. The paper is concluded with a discussion and  detailed review of future directions in section \ref{sec:5}. This article is an expanded accompaniment to the letter \cite{1851681}.

\section{Scattering Amplitudes in Superstring Theory}\label{sec:2}
Scattering amplitudes in superstring perturbation theory are computed according to a prescription that takes into account supersymmetry on the worldsheet. In the case of the bosonic string, amplitudes for $n$ particle scattering at $g$ loops are evaluated by integrating conformal correlators over the moduli space $\mathcal{M}_{g,n}$ of genus $g$ Riemann surfaces with $n$ marked points. When working with superstrings, the correct moduli space to work with is the \emph{supermoduli} space $\mathfrak{M}_{g,n}$. Formally then, the scattering amplitude becomes

\begin{equation}
    \mathcal{A}_{g,n} = \int_{\mathfrak{M}_{g,n}} \langle{ |\delta(H_{A}|B)|^{2}\rangle}_{X,B,C}\times\mathcal{O}_{n}. 
\end{equation}
This formula has been considerably condensed for purposes of readability. To explain some of the notation, $H_{A}$ is a basis of Beltrami superdifferentials on the supermoduli space\footnote{We should note that due to the structure of the moduli space, such a basis is most conveniently defined locally. We will come back to this point when evaluating the integrand on the reduced space $\mathcal{M}_{g,n}$.} while $B$ and $C$ are ghost superfields which (dropping auxiliary fields) are expanded as

\begin{equation}
    B(\theta,z) = b(z) + \theta \beta(z) 
\end{equation}

\begin{equation}
    C(\theta,z) = c(z) + \theta \gamma(z) 
\end{equation}
where $\theta$ and $z$ are local coordinates of the corresponding marked point. $X$ is the target space superfield 

\begin{equation}\label{eq:2.9}
    X^{\mu}(\theta,z,\z) = x^{\mu}(z,\z) + \theta \psi^{\mu}(z) + \overline{\theta}\overline{\psi}^{\mu}(\z).
\end{equation}
Finally, expectation values are evaluated according to the operator product expansions

\begin{equation}
    x^{\mu}(z,\z)x^{\nu}(z',\z') \sim -\eta^{\mu\nu}\alpha\ln|E(z,z')|^{2}
\end{equation}
and

\begin{equation}
    \psi^{\mu}(z)\psi^{\nu}(z') \sim \eta^{\mu\nu}S_{\delta}(z,z').
\end{equation}
Here, $E$ is the prime form on the Riemann surface of genus $g$ while $S_{\delta}$ is the Szego kernel defined for a given spin structure $\delta$. 

The conformal correlator $\mathcal{O}_{n}$ is defined by the correlation function of $n$ vertex operators for the emission of NS bosons. The precise form of these operators will not concern us, since we will employ the chiral splitting theorem \cite{DHoker:1989cxq}, which states that

\begin{equation}\label{eq:2.7}
    \mathcal{O}_{n} = \int_{\mathbb{R}^{10g}}d^{10}p_{I} \bigg|\lab{\exp\left(\frac{i}{\alpha'}\int\chi(z)\psi^{\mu}\partial x^{\mu}(z)\right)\prod_{i}\mathcal{V}(z_{i},\theta_{i},k_{i},\epsilon_{i})}\bigg |^{2}
\end{equation}
where

\begin{equation}
    \mathcal{V}(z_{i},\theta_{i},k_{i},\epsilon_{i}) = \int d\widetilde{\theta}_{i} \exp\left(ik^{\mu}_{i}x^{\mu}_{+}(z_{i})+\frac{2i}{\alpha'}\theta_{i}\widetilde{\theta}_{i}\epsilon^{\mu}_{i}\partial x^{\mu}_{+}(z_{i}) + \theta_{i}k^{\mu}_{i}\psi^{\mu}(z_{i})+\widetilde{\theta}_{i}\epsilon^{\mu}_{i}\psi^{\mu}(z_{i})\right)
\end{equation}
Here, $\chi$ is a gravitino field, which parametrises odd moduli on the supermoduli space while the \emph{chiral} field $x_{+}$ obeys the holomorphic operator product expansion

\begin{equation}
    x^{\mu}_{+}(z)x^{\nu}_{+}(z')\sim -\eta^{\mu\nu} \alpha'\ln(E(z,z')).
\end{equation}

Readers familiar with the chiral splitting theorem will notice that the equation (\ref{eq:2.7}) has not taken into account the loop momentum in the correlator. This is usually incorporated via an insertion of a term of the form

\begin{equation}
    Q(p_{I}) = \exp\left(i\sum_{I}p^{\mu}_{I}\int_{\mathfrak{B}_{I}}\partial x^{\mu}_{+}(z)dz\right),
\end{equation}
which ensures that the zero mode conditions

\begin{equation}
    \int_{\mathfrak{A}_{I}}\partial x^{\mu}_{+}(z)dz = -i\alpha'p^{\mu}_{I}
\end{equation}
are satisfied. Instead of this, we make the equivalent replacement 

\begin{equation}
    \partial x^{\mu}(z) \rightarrow \partial x^{\mu}(z) -i\alpha' p^{\mu}\omega_{I}(z)
\end{equation}
where $\omega_{I}$ are $g$ holomorphic Abelian differentials of the first kind\footnote{The normalization for the zero modes chosen here is the same as that used in \cite{Geyer:2018xwu}, which is somewhat different from the one usually chosen.}. 

We are now tasked with the computation of these chiral correlators. In particular, given a fixed spin structure $\delta$, we are required to evaluate the correlation function

\begin{equation}
 \mathcal{A}_{g,n}[\delta] =    \lab{\prod_{A}\delta(\langle{H_{A}|B\rangle})\exp\left(\frac{i}{\alpha'}\int\chi(z)\psi^{\mu}\partial x^{\mu}(z)\right)\prod_{i}\mathcal{V}(z_{i},\theta_{i},k_{i},\epsilon_{i})}_{B,C,x_{+},\psi}.
\end{equation}

\section{Computing the Chiral Correlator}\label{sec:3}
Before moving to the full calculation of the chiral correlator, we summarise the background of our calculation. First, in the spirit of keeping the analysis completely general, we make no assumptions about the gravitino field $\chi$, only keeping in mind that it is Grassmann valued. Second, although the variables $\widetilde{\theta}$ have appeared here are auxiliary parameters, we will absorb the corresponding integrals into the measure of the supermoduli space, which will later prove computationally convenient. 

To compute the ghost contribution to the chiral correlation function, we need to specify the basis of Beltrami superdifferentials on the moduli space. We will see later that they are defined most conveniently only locally, due to problems involving projectedness of the supermoduli space. Since we defer issues of moving to the reduced space $\mathcal{M}_{g,n}$ to section \ref{sec:4}, we only need to focus on the matter part of the correlation function, which is written as

\begin{equation}
    \mathcal{F}_{g,n} = \lab{\exp\left(\frac{i}{\alpha'}\int\chi(z)\psi^{\mu}\partial x^{\mu}(z)+\sum_{i}\left(ik^{\mu}_{i}x^{\mu}_{+}(z_{i})+\frac{2i}{\alpha'}\theta_{i}\widetilde{\theta}_{i}\epsilon^{\mu}_{i}\partial x^{\mu}_{+}(z_{i}) + \theta_{i}k^{\mu}_{i}\psi^{\mu}(z_{i})+\widetilde{\theta}_{i}\epsilon^{\mu}_{i}\psi^{\mu}(z_{i})\right)\right)}.
\end{equation}
The first step in computing this correlator is to perform the contractions of all the $x_{+}$ fields after carrying out the replacement (\ref{eq:2.9}). Doing this yields

\begin{equation}
\begin{aligned}
    \la\exp\bigg(
    &\alpha'\sum_{i,I}k_{i}\cdot p_{I}\int^{z_{i}}_{P}\omega_{I}(z)dz + \alpha'\sum_{i\neq j}\frac{1}{2}k_{i}\cdot k_{j}\ln (E(z_{i},z_{j})) \\
    &+\int\chi(z)\psi(z)\cdot P(z)dz + \sum_{i}[\theta_{i}k_{i}\cdot\psi(z_{i})+\widetilde{\theta}_{i}\epsilon_{i}\cdot\psi(z_{i})]+ \sum_{i}2\theta_{i}\widetilde{\theta}_{i}\epsilon_{i}\cdot P(z_{i})\\
    &+\sum_{i}\frac{2}{\alpha'}\int \chi(z)\theta_{i}\widetilde{\theta}_{i}\epsilon_{i}\cdot \psi(z) \partial_{z}\partial_{z_{i}}\ln(E(z,z_{i}))dz \\
    &+\sum_{i\neq j}\frac{4}{\alpha'}\theta_{i}\widetilde{\theta}_{i}\theta_{j}\widetilde{\theta}_{j}\epsilon_{i}\cdot \epsilon_{j}\partial_{z_{i}}\partial_{z_{j}}\ln(E(z_{i},z_{j}))\\
    &-\frac{1}{\alpha'}\int\chi(z)\chi(z')\partial_{z}\partial_{z'}\ln(E(z,z'))\psi(z)\psi(z')\bigg )\ra
\end{aligned}
\end{equation}
where

\begin{equation}
    P^{\mu}(z) = \sum_{i}k^{\mu}_{i}\ln(E(z,z_{i})) + \sum_{I}p^{\mu}_{I}\omega_{I}(z). 
\end{equation}
In order to complete the calculation, the expectation value with respect to the spinors $\psi$ must also be carried out. We first note that in the case of odd spin structures, the fermions will have zero modes (denoted by $\psi^{0}$). Keeping the calculation general, upon inclusion of these modes by making the replacement $\psi \rightarrow \psi + \psi^{0}$ we obtain

\begin{equation}\label{eq:3.5}
\begin{aligned}
    \la\exp\bigg(
    &\alpha'\sum_{i,I}k_{i}\cdot p_{I}\int^{z_{i}}_{P}\omega_{I}(z)dz + \alpha'\sum_{i\neq j}\frac{1}{2}k_{i}\cdot k_{j}\ln (E(z_{i},z_{j})) \\
    &+\int\chi(z)\psi(z)\cdot P(z)dz + \sum_{i}[\theta_{i}k_{i}\cdot\psi(z_{i})+\widetilde{\theta}_{i}\epsilon_{i}\cdot\psi(z_{i})]+ \sum_{i}2\theta_{i}\widetilde{\theta}_{i}\epsilon_{i}\cdot P(z_{i})\\
     &+\int\chi(z)\psi^{(0)}\cdot P(z)dz + \sum_{i}[\theta_{i}k_{i}\cdot\psi^{(0)}+\widetilde{\theta}_{i}\epsilon_{i}\cdot\psi^{(0)}]\\
     &+\sum_{i}\frac{2}{\alpha'}\int \chi(z)\theta_{i}\widetilde{\theta}_{i}\epsilon_{i}\cdot \psi^{0} \partial_{z}\partial_{z_{i}}\ln(E(z,z_{i}))dz\\
    &+\sum_{i}\frac{2}{\alpha'}\int \chi(z)\theta_{i}\widetilde{\theta}_{i}\epsilon_{i}\cdot \psi(z) \partial_{z}\partial_{z_{i}}\ln(E(z,z_{i}))dz +\sum_{i\neq j}\frac{4}{\alpha'}\theta_{i}\widetilde{\theta}_{i}\theta_{j}\widetilde{\theta}_{j}\epsilon_{i}\cdot \epsilon_{j}\partial_{z_{i}}\partial_{z_{j}}\ln(E(z_{i},z_{j}))\\
    &-\frac{1}{\alpha'}\int\chi(z)\chi(z')\psi^{0}\psi(z')\partial_{z}\partial_{z'}\ln(E(z,z'))-\frac{1}{\alpha'}\int\chi(z)\chi(z')\psi(z)\psi^{0}\partial_{z}\partial_{z'}\ln(E(z,z'))\\
    &-\frac{1}{\alpha'}\int\chi(z)\chi(z')\partial_{z}\partial_{z'}\ln(E(z,z'))\psi(z)\psi(z')\bigg )\ra
\end{aligned}
\end{equation}
To perform the $\psi$ integral, we note that the last term in the foregoing expression is quadratic, and acts as an effective correction to the kinetic operator $\overline{\partial}$. Indeed, by factorizing out\footnote{The reason for this is that (\ref{eq:2.6}) is the partition function for the fermion with the modified kinetic operator.}

\begin{equation}\label{eq:2.6}
    \lab{\exp\left(\frac{i}{\alpha'}\int\chi(z)\partial x_{+}(z)\cdot\psi(z)dz\right)}
\end{equation}
the correlation function in (\ref{eq:3.5}) can be evaluated simply by replacing the Szego kernel $S_{\delta}(z,z')$ by a an effective Greens function defined by

\begin{equation}
    \left(\delta(z-w)\overline{\partial} - \frac{1}{\alpha'}\int\chi(z)\chi(z')\partial_{z}\partial_{w}\ln(E(z,w))dw\right)\hat{S}(w,z') = \delta(w-z').
\end{equation}
This is an integral equation which can be computed by iteration upon rewriting as

\begin{equation}
    \hat{S}_{\delta}(z,z') = S_{\delta}(z,z') + \frac{1}{\alpha'}\int S_{\delta}(z,w)\chi(w)\partial_{w}\partial_{v}\ln(E(w,v))\chi(v)S_{\delta}(v,z')dwdv.
\end{equation}
With this, we can proceed to carry out the rest of the calculation. Due to the complexity of the final result, it is expressed as a sum

\begin{equation}\label{eq:3.6}
\begin{aligned}
\mathcal{F}_{g,n} = &\lab{\exp\left(i\int\chi(z)\partial x_{+}(z)\cdot\psi(z)dz\right)}\times\\ 
&\exp\left(\mathrm{KN} + H_{0,0} + H_{0,1}+ H'_{0,0} +H_{1,0}+H_{1,1}+H_{2,0}+H_{2,1}\right)
\end{aligned}
\end{equation}
where

\begin{equation}
    \mathrm{KN} = \alpha'\sum_{i,I}k_{i}\cdot p_{I}\int^{z_{i}}_{P}\omega_{I}(z)dz + \alpha'\sum_{i\neq j}\frac{1}{2}k_{i}\cdot k_{j}\ln (E(z_{i},z_{j}))
\end{equation}
is the higher genus analogue of the Koba-Nielsen factor. The term

\begin{equation}
\begin{aligned}
     H_{0,0} = &\int \chi(z)\chi(z')P(z)\cdot P(z')S(z,z')dzdz' + \sum_{i}2\theta_{i}\widetilde{\theta}_{i}\epsilon_{i}\cdot P(z_{i})  \\
     &+2\sum_{i}\left(\int [\chi(z)\theta_{i}P(z)\cdot k_{i}S_{\delta}(z,z_{i})+\chi(z)\widetilde{\theta}_{i}P(z)\cdot\epsilon_{i}S_{\delta}(z,z_{i})]dz\right)\\
     &+\sum_{i\neq j}[\theta_{i}\theta_{j}k_{i}\cdot k_{j}S(z_{i},z_{j})+ \widetilde{\theta_{i}}\widetilde{\theta}_{j}\epsilon_{i}\cdot\epsilon_{j}S(z_{i},z_{j})-2\theta_{i}\widetilde{\theta}_{j}\epsilon_{j}\cdot k_{i}S(z_{j},z_{i})]
\end{aligned}
\end{equation}
is quadratic in all Grassmann coordinates. $H_{0,1}$ is linear in the zero modes of the worldsheet fermions

\begin{equation}
    H_{0,1} = \int\chi(z)\psi^{(0)}\cdot P(z)dz + \sum_{i}[\theta_{i}k_{i}\cdot\psi^{(0)}+\widetilde{\theta}_{i}\epsilon_{i}\cdot\psi^{(0)}].
\end{equation}
The terms cubic and higher in the Grassmann variables are also subleading and above in $\frac{1}{\alpha'}$, defined by

\begin{equation}
    \begin{aligned}
    H_{1,0} = & \frac{2}{\alpha'}\sum_{i}\int[\chi(z)\theta_{i}\widetilde{\theta}_{i}\chi(z')\epsilon_{i}\cdot P(z') \hat{S}(z_{i},z')\partial_{z}\partial_{z_{i}}\ln(E(z,z_{i}))]dzdz'\\
    &+\frac{4}{\alpha'}\sum_{i\neq j}\int[ \chi(z)\theta_{i}\widetilde{\theta}_{i}\theta_{j}\epsilon_{i}\cdot k_{j}\hat{S}(z_{i},z_{j})\partial_{z}\partial_{z_{i}}\ln(E(z,z_{i}))]dzdz'\\
    &+\frac{4}{\alpha'}\sum_{i\neq j}\int[\chi(z)\theta_{i}\widetilde{\theta}_{i}\widetilde{\theta}_{j}\epsilon_{i}\cdot \epsilon_{j}\hat{S}(z_{i},z_{j})\partial_{z}\partial_{z_{i}}\ln(E(z,z_{i}))]dzdz'\\
    &+\frac{4}{\alpha'}\sum_{i\neq j}\left(\theta_{i}\widetilde{\theta}_{i}\theta_{j}\widetilde{\theta}_{j}\epsilon_{i}\cdot\epsilon_{j}\partial_{z_{i}}\partial_{z_{j}}\ln(E(z_{i},z_{j}))\right),
    \end{aligned}
\end{equation}

\begin{equation}
    \begin{aligned}
    H_{1,1} = & \frac{-1}{\alpha'}\int[\chi(z)\chi(z')\chi(w)\psi^{0}\cdot P(w) \hat{S}(z',w)\partial_{z}\partial_{z'}E(z,z')]dzdz'dw\\
    &-\frac{1}{\alpha'}\sum_{i}\int[\chi(z)\chi(z')\theta_{i}\psi^{0}\cdot k_{i}\hat{S}(z',z_{i})\partial_{z}\partial_{z'}E(z,z')]dzdz'\\
    &-\frac{1}{\alpha'}\sum_{i}\int[\chi(z)\chi(z')\widetilde{\theta}_{i}\psi^{0}\cdot \epsilon_{i}\hat{S}(z',z_{i})\partial_{z}\partial_{z'}E(z,z')]dzdz'\\
    &+(z\longleftrightarrow z') + \sum_{i}\frac{2}{\alpha'}\int \chi(z)\theta_{i}\widetilde{\theta}_{i}\epsilon_{i}\cdot \psi^{0} \partial_{z}\partial_{z_{i}}\ln(E(z,z_{i}))dz\\
    \end{aligned}
\end{equation}

\begin{equation}
    \begin{aligned}
    H_{2,0} = \frac{16}{(\alpha')^2}\sum_{i\neq j}\int [\chi(z)\theta_{i}\widetilde{\theta}_{i}\chi(z')\theta_{j}\widetilde{\theta}_{j}\epsilon_{i}\cdot\epsilon_{j}\hat{S}(z,z')\partial_{z}\partial_{z_{i}}\ln(E(z,z_{i}))\partial_{z'}\partial_{z_{j}}\ln(E(z',z_{j}))]dzdz',
    \end{aligned}
\end{equation}
and
\begin{equation}
    \begin{aligned}
    H_{2,1} = & -\frac{2}{(\alpha')^{2}}\sum_{i}\int [\chi(w)\theta_{i}\widetilde{\theta}_{i}\chi(z)\chi(z')\epsilon_{i}\cdot\psi^{0}\hat{S}(w,z')\partial_z\partial_{z'}\ln(E(z,z'))\partial_{w}\partial_{z_{i}}\ln(E(w,z_{i}))]dwdzdz'\\
    &+(z\longleftrightarrow z').
    \end{aligned}
\end{equation}
Finally, $H'_{0,0}$ is obtained by replacing $S$ with $\hat{S}-S$ in the term $H_{0,0}$. Note that this term is subleading and higher order in $\frac{1}{\alpha'}$.

Equation (\ref{eq:3.6}) is the complete expression for the chiral correlation function for the scattering of $n$ NS states at a given spin structure. Modulo contributions from ghosts, this is the final expression for the string integrand. However, it has been defined on the supermoduli space $\mathfrak{M}_{g,n}$. In order to make contact with the ambitwistor string framework, which is defined in terms of integrals over the ordinary moduli space $\mathcal{M}_{g,n}$, we must define how (\ref{eq:3.6}) is expressed on $\mathcal{M}_{g,n}$. 

\section{Reduction from $\mathfrak{M}_{g,n}$ to $\mathcal{M}_{g,n}$}\label{sec:4}

The equation (\ref{eq:3.6}) has been defined in terms of the gravitino field, which must be gauge fixed if we wish to work on the reduced space $\mathcal{M}_{g,n}$ instead of the supermoduli space. This procedure becomes increasingly complicated at higher genus and to an extent remains an open problem, we restrict ourselves to making some general remarks. Henceforth, we work in the limit of $\alpha'\rightarrow \infty$.

Let us start with the simplest case of genus zero. In this case, the supermoduli space $\mathfrak{M}_{0,n}$ is a trivial fibre budle with base $\mathcal{M}_{0,n}$ with $n-2$ fermionic fibres. Indeed, at this order the gravitino field identically vanishes. We have then in the limit $\alpha'\rightarrow \infty$ only the term

\begin{equation}
    H_{0,0}(g=0) = \sum_{i\neq j}\left(\frac{\theta_{i}\theta_{j}k_{i}\cdot k_{j} + \widetilde{\theta}_{i}\widetilde{\theta}_{j}\epsilon_{i}\cdot \epsilon_{j}-2\theta_{i}\widetilde{\theta}_{j}\epsilon_{j}\cdot k_{i}}{z_{i}-z_{j}}\right) + 2\sum_{i}[\theta_{i}\widetilde{\theta}_{j}k_{i}\cdot P(z_{i})].
\end{equation}
The fermionic fibres can be readily integrated out. Note that since there are $n-2$ such fibres, two must be omitted. This choice is arbitrary, and we have the following integrand to be evaluated over $\mathcal{M}_{0,n}$

\begin{equation}
    \int \prod_{i}[d\theta_{k}d\widetilde{\theta}_{k}] \theta_{k}\theta_{\ell}\exp\left(\sum_{i\neq j}\left(\frac{\theta_{i}\theta_{j}k_{i}\cdot k_{j} + \widetilde{\theta}_{i}\widetilde{\theta}_{j}\epsilon_{i}\cdot \epsilon_{j}+2\theta_{i}\widetilde{\theta}_{j}\epsilon_{j}\cdot k_{i}}{z_{i}-z_{j}}\right) + \sum_{i}[\theta_{i}\widetilde{\theta}_{j}k_{i}\cdot P(z_{i})]\right)
\end{equation}
where we have chosen to omit the fermionic directions characterised by $\theta_{k}$ and $\theta_{\ell}$. Indeed, this is precisely the so-called reduced Pfaffian $\mathrm{Pf}_{k\ell}\Psi$ defined in \cite{Cachazo:2013hca}. Note that we have stripped away the Koba-Nielsen term $\mathrm{KN}$ in order to arive at this equivalence\footnote{The tensionless limit of the Koba-Nielsen factor does not impose the saddle points computed by the scattering equations. The inclusion of the Koba-Nielsen factor results in a sum over an infinite number of saddle points, even at tree level, due to localisation on the covering space of the moduli space. See \cite{Mizera:2019vvs} for an illustrative analysis of this phenomenon on at for points for zero genus.}. The ghost contribution at genus zero comes from having to fix three real and two complex superconformal Killing vectors, achieved by $c$ and $\gamma$ insertions (note that the $\gamma$ insertions coincide with the two external states $i,j$ mentioned earlier) - 

\begin{equation}
    \mathcal{Z}^{gh}_{0} = \langle{c(z_{i_1})c(z_{i_2})c(z_{i_3})\delta(\gamma(z_{k}))\delta(\gamma(z_{\ell}))\rangle} = \frac{(z_{i_1}-z_{i_2})(z_{i_2}-z_{i_3})(z_{i_3}-z_{i_1})}{z_k-z_\ell}
\end{equation}

Turning now to the case of genus $1$ with even spin structure, once again $\chi=0$. $H_{0,0}$ reduces to the expression 

\begin{equation}
     H^{\mathrm{even}}_{0,0}(g=1) = \sum_{i}2\theta_{i}\widetilde{\theta}_{i}\epsilon_{i}\cdot P(z_{i})  +\sum_{i\neq j}[\theta_{i}\theta_{j}k_{i}\cdot k_{j}S_{\delta}(z_{i},z_{j})+ \widetilde{\theta_{i}}\widetilde{\theta}_{j}\epsilon_{i}\cdot\epsilon_{j}S_{\delta}(z_{i},z_{j})-2\theta_{i}\widetilde{\theta}_{j}\epsilon_{j}\cdot k_{i}S_{\delta}(z_{j},z_{i})].
\end{equation}
This time, we also need to consider the ghost contribution. For even spin structure, the supermoduli space $\mathfrak{M}_{1,n}$ has $1$ even and $0$ odd moduli and $n-1$ degrees of freedom labelling the marked points. Accordingly, $H_{A}$ is just spanned by one Beltrami differential $\mu$, which parametrises defomations of the complex structure on the torus. Thus, for the ghost contribution we have,

\begin{equation}
    \mathcal{Z}^{gh}_{1}[\delta] = \frac{(\mathrm{Pf}'\overline{\partial})^{10}}{Z^{10}}\langle{(\mu|b)c(z_{i})\rangle}_{bc,\beta\gamma,\delta}.
\end{equation}
Here, the Pfaffian comes from the partition function over the $\psi$ field. The $Z^{10}$ is the chiral scalar partition function coming from the $X$ integration. The ghost contribution in the foregoing is in agreement with the computation in \cite{Adamo:2013tsa}. Such partition functions are computed by bosonisation \cite{Verlinde:1986kw}. 

To compute the matter part of the integrand, we note that there are now $n$ fermionic fibres labelled by Grassmann coordinates on the supermoduli space $\mathfrak{M}_{1,n}$. Accordingly, we are left with the integral

\begin{equation}
\begin{aligned}
    \int \prod_{i}d\theta_{i}d\widetilde{\theta}_{i}\exp\bigg (\widetilde{\theta_{i}}\widetilde{\theta}_{j}\epsilon_{i}\cdot\epsilon_{j}S_{\delta}(z_{i},z_{j})-2\theta_{i}\widetilde{\theta}_{j}\epsilon_{j}\cdot k_{i}S_{\delta}(z_{j},z_{i})]\sum_{i}2\theta_{i}\widetilde{\theta}_{i}\epsilon_{i}\cdot P(z_{i}) \bigg).
\end{aligned}
\end{equation}
This is the Pfaffian $\mathrm{Pf}\Psi_{1,n}$ where

\begin{equation}
  \Psi_{1,n} = \begin{pmatrix}
    \mathbf{A}&-\mathbf{C}^{T}\\
    \mathbf{C}&\mathbf{B}\end{pmatrix}
\end{equation}
where
\begin{equation}
    \mathbf{A}_{ij} = k_{i}\cdot k_{i}S_{\delta}(z_{i},z_{j}),\;\; \mathbf{B}_{ij} = \epsilon_{i}\cdot\epsilon_{j}S_{\delta}(z_{i},z_{j}),\;\;\mathbf{C}_{ij} = \epsilon_{i}\cdot k_{j}S(z_{i},z_{j}),\;\; \mathbf{C}_{ii} = -k_{i}\cdot P(z_{i})
\end{equation}
which matches the chiral correlator computed in \cite{Adamo:2013tsa}.

We turn now to the case of genus one with odd spin structure. In this case, there is one nontrivial fermionic modulus. Projecting along this fibre can be carried out at genus one without any complications by setting

\begin{equation}\label{eq:4.8}
    \chi(z) = \chi_{z_{\alpha_1}}\delta(z-z_{\alpha_1})
\end{equation}
where $\chi_{z_{\alpha_1}}$ is a Grassmann variable to be integrated over. This gauge fixing may first be applied to the ghost system. The even Beltrami differential is then chosen as $\mu$ (parametrising deformations of the complex structure) and the odd one is chosen as $\delta_{z,z_{\alpha_1}}$ to obtain

\begin{equation}
    \mathcal{Z}^{gh}_{1,odd}[\delta] = \frac{(\mathrm{Pf}'\overline{\partial})^{10}}{Z^{10}}\langle{(\mu|b)\delta(\beta(z_{\alpha_1}))c(z_{i})\delta(\gamma(z_{j}))\rangle}_{bc,\beta\gamma,\delta}
\end{equation}
matching \cite{Adamo:2013tsa}. We point out that the $c$ and $\gamma$ insertions corresponding to external states arise due to having to gauge fix superconformal Killing vectors on the torus and can be chosen arbitrarily. We have $n-1$ fermionic directions left to integrate over, which are given by the Grassmann coordinates $\theta_{k}$, where $k$ takes values in $\lbrace{1,2,\dots,i-1,i+1,\dots,n\rbrace}$. Since there are fermionic zero modes, we must carry out integrals over these as well. Using (\ref{eq:4.8}) we obtain

\begin{equation}
    \begin{aligned}
    H^{\mathrm{odd}}_{0,0}(g=1) = & 2\sum_{i}[\chi_{\alpha_1}\theta_{i}P(z_{\alpha_1})\cdot k_{i}S_{\delta}(z_{\alpha_1},z_{i})+\chi_{\alpha_1}\widetilde{\theta}_{i}P(z_{\alpha_1})\cdot \epsilon_{i}S_{\delta}(z_{\alpha_1},z_{i})]+\sum_{i}2\theta_{i}\widetilde{\theta}_{i}\epsilon_{i}\cdot P(z_{i})\\
    &+\sum_{i\neq j}[\theta_{i}\theta_{j}k_{i}\cdot k_{j}S_{\delta}(z_{i},z_{j})+ \widetilde{\theta_{i}}\widetilde{\theta}_{j}\epsilon_{i}\cdot\epsilon_{j}S_{\delta}(z_{i},z_{j})-2\theta_{i}\widetilde{\theta}_{j}\epsilon_{j}\cdot k_{i}S_{\delta}(z_{j},z_{i})].
    \end{aligned}
\end{equation}
and

\begin{equation}
    H^{\mathrm{odd}}_{0,1}(g=1) = \chi_{\alpha_1}\psi^{(0)}\cdot P(z_{\alpha_1}) + \sum_{i}[\theta_{i}k_{i}\cdot\psi^{(0)}+\widetilde{\theta}_{i}\epsilon_{i}\cdot\psi^{(0)}].
\end{equation}
The integrand then becomes

\begin{equation}
    \int d^{10}\psi^{0} d\chi_{\alpha_1}\prod_{k}d\theta_{k}d\widetilde{\theta}_{k}\theta_{i}\exp(H_{0,0}+H_{0,1}).
\end{equation}
This matches the result obtained in \cite{Adamo:2013tsa} for the ambitwistor string. Note that since the integral over $\psi^0$ introduces a ten dimensional Levi-Civita symbol, the above integral is only well defined for five particles or greater. 

At genus two and higher, several complications arise. First, the na\"ive projection used in the genus zero and genus one cases no longer work globally on the supermoduli space. Indeed, even at genus two the moduli space \emph{is} projected, but the calculation is extremely complicated (see the series \cite{DHoker:2001kkt,DHoker:2001qqx,DHoker:2001foj,DHoker:2001jaf,DHoker:2005dys,DHoker:2005vch}). At higher genus however, a projection is not even possible \cite{Donagi:2013dua,Donagi:2014hza}. In other words, the integrand over the supermoduli space cannot be reduced to one on $\mathcal{M}_{g,n}$ by simply integrating out fermionic directions labelled by Grassmann coordinates.

While this is a problem if we wish to obtain an integrand that is valid on \emph{all} of $\mathcal{M}_{g,n}$, there is no obstruction to performing a na\"ive projection if we simply work in a small \emph{neighbourhood} of a point in the moduli space. The supermoduli space can be thought of as a nontrivial fibre bundle over $\mathcal{M}_{g,n}$. As a result, one can always find a basis of fermionic directions to integrate out so long as we work in a small enough open set of $\mathcal{M}_{g,n}$.

Accordingly, let us consider a point $\Sigma\in\mathcal{M}_{g,n}$. There always exists an open set $U_{\Sigma}$ of $\mathcal{M}_{g,n}$ containing $\Sigma$ such that the fermionic directions can be gauge fixed by a choice for the gravitino field of the form

\begin{equation}\label{eq:4.13}
    \chi(z) = \sum_{i=1}^{2g-2}\chi_{\alpha_{i}}\delta(z-z_{\alpha_i}).
\end{equation}
In this neighbourhood, the even Beltrami differentials are simply chosen to be $\lbrace{\mu_{a}\rbrace}$, labelling deformations of the even moduli of $U_{\Sigma}$ (described by a period matrix, say) where $a$ runs over $1,\dots,3g-3$ and the odd differentials are simply $\lbrace{\delta(z-z_{\alpha_i})\rbrace}$. The ghost partition function then takes the familiar form

\begin{equation}
    \mathcal{Z}^{gh}_{g}[\delta] = \frac{(\mathrm{Pf}'\overline{\partial})^{10}}{Z^{10}}\lab{\prod_{a}(\mu_{a}|b)\prod_{i}\delta(\beta(z_{\alpha_i}))}_{bc,\beta\gamma,\delta}.
\end{equation}
This compares favourably to $\mathcal{Z}_{gh}[\delta]$ found in \cite{Geyer:2018xwu}. Using this parametrisation of the fermionic fibres, we have for the integrand in the case of even spin structure

\begin{equation}\label{eq:4.15}
    \int \prod_{i}d\chi_{\alpha_i}\prod_{k}d\theta_{k}d\widetilde{\theta}_{k}\exp(H_{0,0})
\end{equation}
where

\begin{equation}
    \begin{aligned}
    H_{0,0} = &\sum_{\alpha_i,\alpha_j}\chi_{\alpha_i}\chi_{\alpha_j}P(z_{\alpha_i})\cdot P(z_{\alpha_{j}})S_{\delta}(z_{\alpha_{i}},z_{\alpha_{j}}) + \sum_{i}2\theta_{i}\widetilde{\theta}_{i}\epsilon_{i}\cdot P(z_{i})  \\
     &+2\sum_{z_{\alpha_i},i} [\chi_{\alpha_i}\theta_{i}P(z_{\alpha_i})\cdot k_{i}S_{\delta}(z_{\alpha_i},z_{i})+\chi_{\alpha_i}\widetilde{\theta}_{i}P(z_{\alpha_i})\cdot\epsilon_{i}S_{\delta}(z_{\alpha_i},z_{i})]\\
     &+\sum_{i\neq j}[\theta_{i}\theta_{j}k_{i}\cdot k_{j}S_{\delta}(z_{i},z_{j})+ \widetilde{\theta_{i}}\widetilde{\theta}_{j}\epsilon_{i}\cdot\epsilon_{j}S_{\delta}(z_{i},z_{j})-2\theta_{i}\widetilde{\theta}_{j}\epsilon_{j}\cdot k_{i}S_{\delta}(z_{j},z_{i})]
    \end{aligned}
\end{equation}
Indeed, (\ref{eq:4.15}) takes the compact form $\mathrm{Pf}'\Psi_{g,n}$ where

\begin{equation}
  \Psi_{g,n} = \begin{pmatrix}
    \mathbf{A}&-\mathbf{C}^{T}\\
    \mathbf{C}&\mathbf{B}\end{pmatrix}
\end{equation}
and

\begin{equation}
\begin{aligned}
    &\mathbf{A}_{\alpha_i\alpha_j} = P(z_{\alpha_i})\cdot P(z_{\alpha_{j}})S_{\delta}(z_{\alpha_{i}},z_{\alpha_{j}})\;\; \mathbf{A}_{\alpha_i i} = P(z_{\alpha_i})\cdot k_{i}S_{\delta}(z_{\alpha_i},z_{i}), \;\; \mathbf{C}_{\alpha_i j} = P(z_{\alpha_i})\cdot \epsilon_{i}S_{\delta}(z_{i},z_{\alpha_i})\\
    & \mathbf{A}_{ij} = k_{i}\cdot k_{i}S_{\delta}(z_{i},z_{j}),\;\; \mathbf{B}_{ij} = \epsilon_{i}\cdot\epsilon_{j}S_{\delta}(z_{i},z_{j}),\;\;\mathbf{C}_{ij} = \epsilon_{i}\cdot k_{j}S(z_{i},z_{j}),\;\; \mathbf{C}_{ii} = -k_{i}\cdot P(z_{i}).
\end{aligned}
\end{equation}
This is identified with the chiral correlation function computed for even spin structure in \cite{Geyer:2018xwu}. Parenthetically, for odd spin structures (\ref{eq:4.15}) is modified to

\begin{equation}
     \int d^{10}\psi^{(0)}\prod_{i}d\chi_{\alpha_i}\prod_{k}d\theta_{k}d\widetilde{\theta}_{k}\exp(H_{0,0}+H_{0,1})
\end{equation}
where

\begin{equation}
    H_{0,1}  = \sum_{\alpha_i}\chi_{\alpha_i}\psi^{(0)}\cdot P(z_{\alpha_i}) + \sum_{i}[\theta_{i}k_{i}\cdot\psi^{(0)}+\widetilde{\theta}_{i}\epsilon_{i}\cdot\psi^{(0)}].
\end{equation}
Once again, this is precisely the chiral correlation function computed by the ambitwistor string. We remark here that in analysing the ambitwistor string the main interest is in scattering amplitudes in below $10$ dimensions. Accordingly, the contributions from odd spin structures vanish identically. We have nevertheless chosen to include a discussion of them for the sake of completeness.

To readers familiar with the results in \cite{Geyer:2018xwu}, it may appear inconsistent that the authors have made the choice (\ref{eq:4.13}) to define, at least formally, an integrand on the whole of $\mathcal{M}_{g,n}$, given the obstruction to projection that we have noted. To clarify this discrepancy, we need to understand what exactly the issue of non-projectedness entails when we are working on the reduced space $\mathcal{M}_{g,n}$.

Projecting down to $\mathcal{M}_{g,n}$ (keeping in mind that we must work in a small open set thereof) results in the insertion of the so-called picture changing operators $i\frac{\delta(\beta(z_{\alpha_i}))}{\alpha'}\psi(z_{\alpha_i})\cdot\partial x_{+}(z_{\alpha_i})$ into the superstring correlation function. Permission to project down to the ordinary moduli space globally is equivalent to being able to choose the points $z_{\alpha_i}$ consistently and uniquely over the entire moduli space. Any projection must therefore be at best a local prescription for determining a the points $z_{\alpha_i}$ by a function $f_{i}(m)$ of the local bosonic moduli. However, these functions must be chosen avoiding a locus of so-called spurious singularities - which is known to be of complex codimension $1$ \cite{Sen:2015hia}. Spurious singularities are of four kinds, which we now review.
\begin{enumerate}
\item Spurious singularities of the first kind arise when $f_{\alpha_i}(m) = f_{\alpha_j}(m)$ for some $\alpha_i$ and $\alpha_j$. In such a case, two of the picture changing operators approach one another, leading to a divergence.

\item Spurious singularities of the second kind arise when one the of PCO insertions approaches a marked point $z_{i}$, leading to a divergence. 

\item Spurious singularities of the third kind are more subtle. They arise when the choice (\ref{eq:4.13}) does not furnish a complete basis for the fermionic moduli of the supermoduli space. Indeed, it should be thought of as a singularity due to a poor choice of gauge. This happens when there exists a nontrivial solution $y$ on some $\Sigma(m)\in\mathcal{M}_{g,n}$ to the equation

\begin{equation}
    \overline{\partial} y = \sum_{\alpha_i}e_{\alpha_i}\delta(z-z_{\alpha_i})
\end{equation}
where not all $e_{\alpha_i}$ vanish. It is to avoid such singularities that the PCOs must be chosen to vary holomorphically with the moduli.

\item Spurious singularities of the fourth kind are genuine divergences. They are independent of the choice of marked points and PCO insertions. To understand these, we need to go back to the ghost partition function. Upon computation by bosonisation, the ghost partition function contains $\vartheta$ functions in the denominator. They can be shown to only depend on the even moduli of $\mathcal{M}_{g,n}$. At points where they vanish, we encounter the fourth kind of spurious divergence.
\end{enumerate}

In dealing with the spurious singularities of the first and second kinds, it was observed in \cite{Geyer:2018xwu} that when these singularities are approached, the integrand of the ambitwistor string vanishes on the support of the scattering equations. Liuoville's theorem then ensures independence from the choice of PCO insertions. In that work however, the spurious singularities of the third and fourth kind were not considered.

To understand why they did not lead to any inconsistencies, we note that at least at genus two, the ambitwistor string was shown to localise on a discrete set of $(n-3+2g)!$ points on the moduli space $\mathcal{M}_{2,n}$, first by degenerating two $\mathfrak{A}$ cycles and then by evaluating the integrand on the scattering equations on the binodal sphere. Indeed, the spurious singularities of the third kind form a set of real codimension $2$, and for a specific $\Sigma\in\mathcal{M}_{g,n}$ entail preventing one from inserting PCOs at a finite set of points. Consequently, modulo what ends up being a finite set, PCOs may be inserted anywhere else, causing no problems thereafter.

The case of the last kind of singularity is more subtle. In this case, the authors of \cite{Geyer:2018xwu} have essentially established that no such singularity arises when the integrand is restricted to nonseparating degenerations of $\mathcal{M}_{2,n}$. This was done by direct evaluation of the ghost partition function on this subspace.

In order to see how this works at higher genus, there are two distinct facts to be verified. First, we would need to be certain that the ambitwistor string localises on a finite set of points on $\mathcal{M}_{g,n}$. Second, we would need to ensure that on the support of this finite set (or within small neigbourhoods of the points therein), the theta functions in the definition of the ghost partition functions do not vanish. This analysis is beyond the scope of the present work. Accordingly, the relation between the tensionless limit of the  chiral string correlator and the ambitwistor correlator should be regarded as a being true in small neighbourhoods, not globally.

As an aside, circumventing spurious singularities in full superstring perturbation theory is exceedingly difficult. To ensure that the PCOs are inserted in a manner consistent with unitarity and that spurious singularities are avoided, an integration cycle $\Gamma_{V}\in\mathcal{M}_{g,n}$ was constructed in \cite{Sen:2014pia,Sen:2015hia}. Notably, since there is generically no global choice of PCOs, gluing local presentations of the integrand has to be done carefully, while integrating over a cycle which ensures that spurious singularities are not encountered.  We won't consider this any further here, but a thorough understanding of this procedure for the ambitwistor string may be relevant at higher genus order.

\section{Discussion}\label{sec:5}
In this paper, we have employed the chiral splitting theorem to compute the chiral superstring integrands for $n$ NS states at finite $\alpha'$. Since these integrands are formally defined on the supermoduli space $\mathfrak{M}_{g,n}$, we had to move to the ordinary moduli space to make a comparison with the chiral half integrands in the ambitwistor string.

The reduction procedure is complicated at higher genus due to the non-projectedness of the supermoduli space. Accordingly, we had to perform the reduction procedure only locally, where we are always guaranteed to have a consistent projection. Once this reduction was performed, we observed that in the limit of vanishing tension, the chiral integrands of the superstring matched those computed by the ambitwistor string. With this result, we have on hand at least formal expressions that provide the right generalisation of the ambitwistor integrands for finite $\alpha'$. Let us now review some further issues that may be serve as interesting problems for future research.

\vspace{0.5cm}
\paragraph{\textbf{Intersection Numbers and the Field Theory Limit}}
The heuristic derivation of the ambitwistor string as the $\alpha'\rightarrow 0$ limit of the RNS superstring in \cite{Mason:2013sva} together with the results of this paper suggests that a further analysis of the relationship between the tensionless and infinite tension limits are required. One possible manner in which such a study might be carried out is the framework of \emph{twisted intersection theory}.

The equivalence between the infinite tension field theory and tensionless limits of superstrings at tree level was established by the formalism of Cachazo, He and Yuan. A full understanding of why this is the case was supplied in \cite{Mizera:2017cqs,Mizera:2017rqa}, where it was observed that the right way of thinking about the localisation formula (\ref{eq:1.2}) was in terms of so called \emph{intersection numbers}. Intersection numbers are topological invariants defined by a cohomology that arises out of a twist 
\begin{equation}
    \nabla_{\omega} = d + d\omega\wedge
\end{equation} 
where $\omega$ is a Morse function whose derivative coincides with the scattering equations. When viewed as intersection numbers, the amplitudes computed by the CHY formula are simply intersection numbers of logarithmic representatives of the twisted cohomology group, which can be shown to always be $\alpha'$ independent.   

Establishing the equivalence of the tensionless and infinite tension limits would require a proof that the chiral integrand derived in this paper can always be recast in logarithmic form (at least on the nodal Riemann sphere).

\vspace{0.5cm}
\paragraph{\textbf{Implications for the Duality of Colour and Kinematics}}
The result of section \ref{sec:4} can be summarised as establishing that for a given point $\Sigma\in\mathcal{M}_{g,n}$ there is a neighbourhood $U_{\Sigma}$ where the chiral superstring integrand admits an expansion of the form

\begin{equation}\label{eq:5.1}
    \mathcal{I}(\alpha') = \sum_{\delta}Z^{gh}_{g}[\delta]\left(\mathrm{Pf}\Psi_{g,n}+ \mathcal{O}\left(\frac{1}{\alpha'}\right)\right)
\end{equation}
where the first term yields the correct half integrand for $n$ particle NS states in the ambitwistor string. To compute scattering amplitudes of gravitons, two copies of this half integrand are integrated over $\mathcal{M}_{g,n}$ on the support of the scattering equations. We recall that the scattering equations are obtained by demanding that $P(z)\cdot P(z)$ vanish pointwise on the moduli space. Since the moduli space is $3g-3+n$ dimensional, this constraint can be imposed by setting to zero $\langle{\mu_{i}P^2\rangle}$ for a $n-3+3g$ dimensional basis of Beltrami differentials. At least till genus two, it is known that the ambitwistor string integral localises on a collection of one particle irreducible trivalent graphs in the form of a Feynman expansion,

\begin{equation}\label{eq:5.2}
    \mathcal{A}_{g,n} = \sum_{v_{\Gamma}\in \Gamma_{g}}\frac{\mathrm{Res}_{v_{\Gamma}}(\mathcal{I}_{L})\mathrm{Res}_{v_{\Gamma}}(\mathcal{I}_{R})}{\prod_{e\in \Gamma_{g}}p_{e}}
\end{equation}
where the sum runs over graphs contained in $\Gamma_{g}$ - the set of trivalent graphs with $g$ loops. The denominators are products over momentum transfers, where each loop is associated to a separate loop momentum. Note that the $\mathcal{I}$ is the $\alpha'\rightarrow \infty$ limit of (\ref{eq:5.1}).

It was argued in \cite{Mizera:2019blq} using twisted intersection numbers that under conditions where an expansion of the form (\ref{eq:5.2}) holds, the numerators automatically satisfy the duality between colour and kinematics due to Bern, Carrasco and Johansson \cite{Bern:2005hs} when the residues are evaluated on degenerate Riemann spheres. However, this requires taking the finite $\alpha'$ expressions for $\mathcal{I}(\alpha')$ and removing $\alpha'$ dependence. While this was done in \cite{Mizera:2019blq} by finding the correct representative in twisted cohomology, the same result can be achieved in a far more pedestrian fashion by repeated integrations by parts \cite{Bjerrum-Bohr:2014qwa}. To review this argument, note that at genus zero the full $\alpha'$ expansion of the integrand is as follows

\begin{equation}
    \mathrm{Pf}\Psi_{[12]} + \sum_{k=1}^{\frac{n-1}{2}}\frac{1}{(\alpha')^{k}}\prod_{\mathrm{distinct\;pairs}\lbrace{i_{\ell},j_{\ell}\rbrace}}\left(\frac{\epsilon_{i_{\ell}}\cdot\epsilon_{j_{\ell}}}{(z_{i_{\ell}}-z_{j_{\ell}})^{2}}\right)\mathrm{Pf}\Psi^{[i_{1}j_{1}\dots i_{k}j_{k}]}_{12}
\end{equation}
where in the subleading terms columns and rows labelled by $i_{\ell}$, $j_{\ell}$, $n+i_{\ell}$ and $n+j_{\ell}$ are removed and $\ell$ runs from $1$ through $k$. This removal makes the Pfaffian terms independent of the corresponding marked points. In analogy to the removal of second derivative due to Bern and Kosower, repeated integrations by parts acting on the global Koba-Nielsen factors yield terms proportial to the scattering equations and independent of $\alpha'$, after which the $\alpha'\rightarrow 0$ limit is taken instead, to recover the field theory limit.

This method manifestly applies in the genus zero and genus one cases. To complete a proof of the colour-kinematics duality this way, we would have to establish a similar equivalence between the $\alpha'\rightarrow 0$ and $\alpha'\rightarrow\infty$ limits of $\mathcal{I}(\alpha')$ at higher genus as well. This likely requires a more careful study of the contour of integration at higher genus.

\vspace{0.5cm}
\paragraph{\textbf{Ramond Sector Vertex Operators}}

In this paper, we have only considered the chirally split amplitudes for external states of NS type. Ramond sector vertex operators are defined in terms of the spin field $S_{\alpha}$, which has nontrivial operator product expansions with itself as well as with the worldsheet fermions. 

Since for the most part, the literature has only dealt with external NS states in the context of the ambitwistor string, we have not taken into account Ramond states in this work. Accordingly, the complete calculation of the chiral half integrand for arbitrary numbers of NS and R states in the superstring and the comparison of its tensionless limit to the corresponding correlation function of the ambitwistor string remains an open problem. 

On this note, it is worth pointing out that although for purposes of completeness we have computed the chiral integrands for odd and even spin structures, for the reason that the ambitwistor string has been used to obtain scattering amplitudes in dimensions less that $10$, the odd spin structure contributions have not been studied in the literature to as great an extent as the even spin structures. Indeed, inclusion of odd spin structures demands us to consider intermediate states that are of Ramond type. This can be circumvented by restricting the spin sum to a particular class of even spin structures. See \cite{Geyer:2018xwu} for details at genus two. 

\vspace{0.5cm}
\paragraph{\textbf{More General Backgrounds}}
Most formulae of CHY type have focussed on scattering amplitudes in flat space backgrounds. Conveniently, the chiral splitting theorem holds true in flat space background, making it possible for us to carry out the analysis done in this paper.

Left out however is the study of scattering amplitudes in more generic backgrounds. Some progress was made in \cite{Eberhardt:2020ewh,Roehrig:2020kck}, where the authors derived the CHY formula for scalar conformal correlators in anti de Sitter space. Here, the scattering equations are modified to take the form

\begin{equation}
    E^{\mathrm{AdS}}_{i} = \sum_{j\neq i}\frac{D_{i}\cdot D_{i}}{z_{i}-z_{j}}
\end{equation}
where the $D_{i}$ are generators of the conformal group in AdS space. Indeed, the scattering equations in this case are operator valued, unlike the algebraic equations obtained in the flat space case. A similar formalism was developed in \cite{Casali:2020uvr} to describe Mellin space scattering amplitudes by CHY formulae. Generalising the analysis performed in this paper to generic backgrounds will require a consistent formalism of string perturbation theory to more general spacetimes and corresponding generalisations of the chiral splitting theorem.

\section*{Acknowledgements}
I thank Jacob Bourjaily for discussions at various stages of the project and going over the draft. I have been fortunate to have had helpful correspondence with Yvonne Geyer, Alok Laddha, Sebastian Mizera, Ricardo Monteiro, Seyed Faroogh Moosavian, Oliver Schlotterer and Edward Witten. Early conversations on chiral splitting with Seyed Faroogh Moosavian were especially crucial for my consideration of this problem. This project has been supported by an ERC Starting Grant (No. 757978) and a grant from the Villum Fonden (No. 15369).

\appendix

\section{Some Facts about the Supermoduli Space $\mathfrak{M}_{g,n}$}
We review some basic facts regarding the supermoduli space which we have used in section \ref{sec:4}. We refer readers interested in more details about the supermoduli space and superstring perturbation theory to \cite{Witten:2012bg,Witten:2012ga,Witten:2012bh,Witten:2013tpa,Witten:2013pra,Moosavian:2019pmd}. We note that by $\mathfrak{M}_{g,n}$, all the puctures are taken to be of NS type\footnote{Including Ramond punctures turns out to be considerably more complicated. See \cite{ott2020supermoduli} for a discussion at genus zero.}. 

The supermoduli space $\mathfrak{M}_{g,n}$ is the moduli space of genus $g$ super Riemann surfaces with $n$ marked points. This is easiest to understand at genus zero, where it can be defined as the quotient of the configuration space of $n$ points on $\mathbb{CP}^{1|1}$ by the group $\mathrm{OSp}(2|1)$

\begin{equation}
    \mathfrak{M}_{0,n} = \mathrm{Conf}\left(\mathbb{CP}^{1|1},n\right)/\mathrm{OSp}(2|1)
\end{equation}
which is equivalent to $\mathbb{CP}^{n-3|n-2}$ modulo coincident points. Accordingly, $\mathfrak{M}_{0,n}$ is of complex dimension $(n-3|n-2)$. This is generalised at higher genus to $(n+3g-3|n+2g-2)$. There is however a subtlety at genus one. For genus one in the case of even spin structure, there are no nontrivial fermionic moduli. Hence, the odd directions are parametrised entirely by Grassmann coordinates. For odd spin structure however, there is one fermionic modulus. Accordingly, 
$n-1$ odd directions are labelled by Grassmann coordinates, while the odd modulus is carried by a gravitino field $\chi$.
This is summarised by 

\begin{equation}
    \begin{aligned}
    \mathrm{dim}\left(\mathfrak{M}_{0}\right)  = &(0|0)\\
    \mathrm{dim}\left(\mathfrak{M}_{1}\right)_{\mathrm{even}}  = &(1|0)\\
    \mathrm{dim}\left(\mathfrak{M}_{1}\right)_{\mathrm{odd}}  = &(1|1)\\
    \mathrm{dim}\left(\mathfrak{M}_{g}\right)  = &(3g-3|2g-2).\\
    \end{aligned}
\end{equation}
Up to and including at least genus 2, the supermoduli space is \emph{projected}, meaning that a consistent global choice of bosonic coordinates can be made such that the fermionic directions can be integrated out to recover expressions valid on all of $\mathcal{M}_{g,n}$. This is known to fail at least starting at genus 5 \cite{Donagi:2013dua,Donagi:2014hza} for $\mathfrak{M}_{g}$, while projectedness remains an open questions in the presence of punctures. Accordingly, in order to bring formulae valid on the supermoduli space to ones on $\mathcal{M}_{g,n}$, it is best to work on open sets $\mathcal{U}\in\mathcal{M}_{g,n}$ onto which, when small enough, one can perform projections. 

For suitably small open sets, the supermoduli space (when regarded as a nontrivial bundle over $\mathcal{M}_{g,n}$) can be projected by simply describing the moduli space in terms of a two-dimensional metric $g_{mn}(m^{a})$ dependding only on the bosonic moduli and a gravitino slice of the form (\ref{eq:4.13}). As mentioned earlier, this has to be done taking into account the issue of spurious divergences. Patching together local descriptions of this sort poses a major challenge in full superstring theory, where use must be made of the procedure of vertical integration \cite{Sen:2014pia,Sen:2015hia}. 
\bibliographystyle{JHEP}
\bibliography{main}
\end{document}